\documentclass[journal=jacsat,manuscript=article]{achemso}
\setkeys{acs}{keywords = true}

\usepackage{chemformula} 

\author{Alok Pokharel}
\affiliation{Univ. Lille, CNRS, Centrale Lille, Univ. Polytechnique Hauts-de-France, UMR 8520 - IEMN, F-59000 Lille, France }
\author{Hao Xu}
\affiliation{Univ. Lille, CNRS, Centrale Lille, Univ. Polytechnique Hauts-de-France, UMR 8520 - IEMN, F-59000 Lille, France }
\author{Srisaran Venkatachalam}
\affiliation{Univ. Lille, CNRS, Centrale Lille, Univ. Polytechnique Hauts-de-France, UMR 8520 - IEMN, F-59000 Lille, France }
\author{Eddy Collin}
\affiliation{Univ. Grenoble Alpes, Institut NEEL - CNRS UPR2940, 25 rue des Martyrs, BP 166, 38042 Grenoble Cedex 9, France}
\author{Xin Zhou}
\email{Corresponding Author:  xin.zhou@cnrs.fr}
\affiliation{Univ. Lille, CNRS, Centrale Lille, Univ. Polytechnique Hauts-de-France, UMR 8520 - IEMN, F-59000 Lille, France }

\title[An \textsf{achemso} demo]
  {Capacitively coupled distinct mechanical resonators for room temperature phonon-cavity electromechanics}

\keywords{phonon-cavity, coupled mechanical resonators, membrane, two-tone, interference}

\begin{document}


\begin{abstract}
Coupled electromechanical resonators that can be independently driven/detected and easily integrated with external circuits are essential for exploring mechanical modes based signal processing. Here, we present a room temperature phonon-cavity electromechanical system, consisting of two distinct resonators: a silicon nitride electromechanical drum capacitively coupled to an aluminum one. We demonstrate electromechanically induced transparency and amplification in a two-tone driving scheme and observe the phonon-cavity force affecting the mechanical damping rates of both movable objects. We also develop an analytical model based on linearly coupled motion equations, which captures the optomechanical features in the classical limit and enables to fit quantitatively our measurements. Our results open up new possibilities in the study of phonon-cavity based signal processing in the classical and potentially in the future in the quantum regimes. 
\end{abstract}

\section{Introduction}
Micro- and nano-electromechanical systems (MEMS and NEMS), allowing mechanical displacements to couple with electrical and optical signals, have been studied for various applications and fundamental research \cite{bachtold2022mesoscopic}. The specific features of tiny scale and high quality factor are attractive for sensing applications \cite{spletzer2006ultrasensitive, spletzer2008highly}. Their intrinsic nonlinearity and mechanical transduction design are explored for developing logic gates \cite{guerra2010noise,mahboob2011interconnect},  radio frequency (RF) amplifiers \cite{karabalin2011signal} and memory nodes \cite{mahboob2008bit}. In recent years, great efforts have been made for investigating mode coupling, which exists  between different mechanical modes in a single system and between different resonators. Not only because the coupled mechanical modes enhance the sensitivity in detection \cite{spletzer2008highly}, but because they also give access to  energy exchange between two different mechanical modes \cite{okamoto2013coherent, faust2012nonadiabatic}, which enables e.g. to transmit/filter information in different frequency bands\cite{bannon2000high}. Besides, mode coupling offers an opportunity to simulate quantum systems by emulating a classical two-level system, and opens up new applications in quantum information processing by using NEMS/MEMS integrated optomechanical systems \cite{faust2013coherent, fan2015cascaded, shahidani2013control}. 

Inspired by recent achievements in optomechanics, the concept of phonon-cavity has been widely exploited in studies of mechanical mode coupling \cite{mahboob2012phonon, sun2016correlated}. In order to generate phonon-phonon interactions, the typical method is to implement the mechanical mode having the higher resonance frequency $\Omega_1$ in the coupled system as a ``phonon cavity" (in analogy with an optical cavity), and to sideband pump it at the frequency $\sim \Omega_1 \pm \Omega_2$, where $\Omega_2$ is the resonance frequency of the other mode. Rich physics from optomechanics has been inherited in phonon-cavity electromechancial systems, such as the optical spring effect or the amplification/de-amplification of the mechanical damping rate, which can lead to self-sustained oscillation states generated by the cavity force  \cite{sun2016correlated, zeng2021strong, cattiaux2020beyond}. These coupled mechanical modes also enrich existing optomechanical functions. It has been proved that an optomechanical system,  consisting of coupled mechanical modes, exhibits additional flexibility in adjusting the optomechanical transparency window for signal processing and increasing information storage time \cite{fan2015cascaded, shahidani2013control}. In most phonon-cavity schemes, the mechanical mode coupling is created between different mechanical modes with a single resonator by means of an intrinsic nonlinearity, or between different resonators by using physical connections transmitting a displacement-induced tension \cite{sun2016correlated, okamoto2013coherent, spletzer2006ultrasensitive, mahboob2015dispersive, karabalin2009nonlinear, de2016tunable}.  However, mechanical coupling design yields implementation complexities in optimization of the coupling between distinct resonators, and poses a challenge when electromechanical devices with higher resonance frequencies and flexible frequency tunability are required. Compared to these more common mechanical coupling designs, capacitive coupling schemes bring a convenient capability to couple distributed electromechanical resonators by giving  them the capability to directly exchange information with each other via cavity phonons  \cite{fan2015cascaded, shahidani2013control}. However, it is still challenging to realize directly coupled mechanical resonators via a capacitive coupling scheme \cite{siskins2021tunable, huang2013demonstration}.

In this work, we present mechanical motion transduction between two  capacitively coupled, and distinct electromechanical resonators, consisting of an Al drum and a SiN drum. Both resonators can be driven and detected independently. In a two-tone driving scheme, we explore phonon-cavity electromechanics based on a simple theoretical model which is analogue to microwave optomechanics, and experimentally demonstrate electromechanically induced transparency and amplification at room temperature, by pumping the phonon-cavity at its sideband to generate phonon-phonon interactions. Modulations of the mechanical damping rate (with respect to the applied driving tone) have been observed in both coupled drum resonators, and exhibit the trend expected by the theoretical model. These results indicate that this new type of device design could serve for phonon-based information processing in both classical and quantum regimes, such as information storage and tunable filter \cite{fan2015cascaded, shahidani2013control}.
 

The device structure investigated in this work consists of two distinct  electromechanical resonators. One of them is a silicon nitride (SiN) drum, $\approx$ 80 nm in thickness, covered with an aluminum (Al) thin film $\sim$ 25 nm in thickness, which has been fabricated from a silicon substrate covered with a stoichiometric silicon nitride thin film, $\sim$ 1 GPa tensile stress. The SiN drum measured in this work is $\approx$ 36 $\mu$m in diameter. The other coupled resonator is an Al drum having a diameter of $\approx$ 40 $\mu$m, suspended on top of the SiN drum. Details of device fabrication have been reported in our previous work \cite{zhou2021high}.
\begin{figure}
  \includegraphics[width=0.98\textwidth]{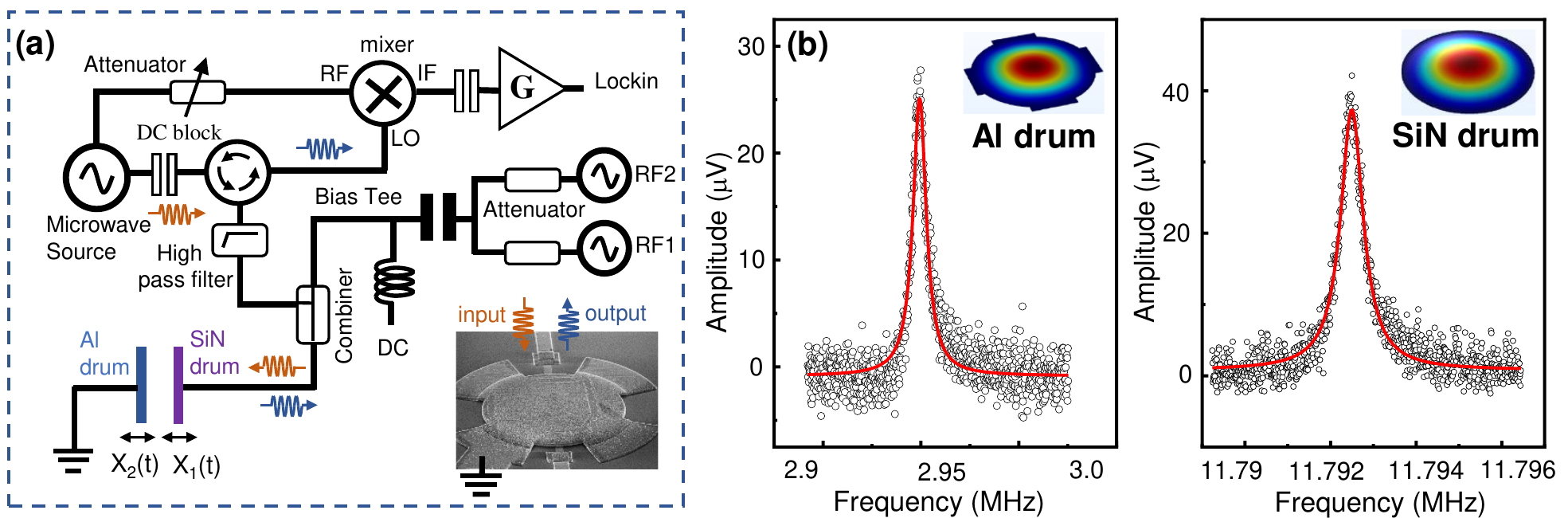}
  \caption{(a) Schematic diagram of the measurement setup. Both electromechanical resonators are driven by low frequency signals combined with RF and \textit{dc} signals. The mechanical displacement is imprinted in the reflected microwave signals and read out by a lockin amplifier through frequency down-conversion \cite{zhou2021high}. The inset shows a scanning electron microscope (SEM) image of an Al drum resonator, which is suspended on top of a SiN drum covered with an Al thin film. The Al drum  is designed to have a ``X"-shaped clamping structure covering $\sim$50\text{\%} of the circumference. (b) Linear resonance response of the Al drum resonator (left) and SiN drum resonator (right), which are measured at $V_{dc}$=2 V, $V_{ac}$=2 m$V_p$ and 0.2 m$V_p$ respectively. The inset figures show the corresponding mechanical mode shapes obtained from finite element simulations.}
  \label{sch:setup}
\end{figure}
The measurement setup is schematically depicted in Figure \ref{sch:setup} (a). Mechanical motions of both electromechanical resonators can be independently excited by passing RF signals $V_{ac}$ combined with \textit{dc} voltages $V_{dc}$ to generate an electrostatic driving force. All the  measurements are performed at room temperature, under vacuum ($\sim 10^{-6}$ mbar) to minimize air damping. Figure \ref{sch:setup} (b) shows linear responses of both Al drum and SiN drum resonators, for their fundamental modes.  For the Al drum resonator, the resonance frequency is $\Omega_{Al}$/(2$\pi$) $\approx$ 2.95 MHz with a quality factor $Q_{Al}\approx$ 358. A finite element simulation is consistent with a low tensile stress in the Al film $\sim$40 MPa, which could be induced during the electron beam evaporation process. The SiN drum vibrates at frequency $\Omega_{SiN}$/(2$\pi$) $\approx$ 11.792 MHz with a quality factor $Q_{SiN}\approx$ 1.8$\times 10^4$, far from the resonance range of the Al drum resonator. In the experiment, we take the SiN resonator as the phonon-cavity. It actually presents an energy leaking rate which is much smaller than that of the coupled Al drum, contrary to standard optomechanical systems. 

\subsection{RESULTS AND DISCUSSIONS}

\textbf{Electromechanical capacitive coupling model.} The whole device structure can be viewed as a parallel plate capacitor $C_g\left(X_1, X_2\right)$, where each plate is a membrane drum resonator, as shown in Figure \ref{sch:setup} (a). The mechanical displacement of each membrane is described by $X_1(t)$ and $X_2(t)$ resonating at the frequency $\Omega_1$ and $\Omega_2$ respectively far from each other,  with $\Omega_1 > \Omega_2$. Driven by the electrostatic force $F_{1,2}(t)$ = $\frac{\left[ V_{dc}+V_{ac}\left( t \right) \right]^2}{2}\frac{\partial }{\partial X_{1,2}}C_g\left[ X_1\left( t\right), X_2\left( t\right) \right]$, we therefore model these two capacitively coupled drums in the linear response regime via the following coupled equations of motion for the displacements $X_1(t)$ and $X_2(t)$, 
\begin{equation}
\begin{aligned}
  \ddot{X}_1 + \gamma_1\dot{X_1} +\Omega^2_{1} X_1 &=  \frac{V_{ac}V_{dc}}{m_1 \, d} C_{g0}\left[1-\frac{2(X_2-X_1)}{d}\right], \\
  \ddot{X}_2 + \gamma_2\dot{X_2} +\Omega^2_{2} X_2 &=  \frac{V_{ac}V_{dc}}{m_2 \, d} C_{g0}\left[-1+\frac{2(X_2-X_1)}{d}\right].
  \label{eqn:motion}
\end{aligned}
\end{equation}
Here,  $\gamma_{1, 2}$ are the mechanical damping rates and $m_{1,2}$ are the effective masses of each drum. The driving force, acting on a simple parallel plate capacitor, is truncated at the second order in a Taylor expansion, and $C_{g0}$ is the initial capacitance between the two membranes separated by a distance $d$. In Eq.\ref{eqn:motion}, the approximation $2V_{dc}V_{ac}+V_{ac}^2 \approx 2V_{dc}V_{ac}$ has been made by considering the typical situation encountered in measurements: $V_{dc} \gg \vert V_{ac} \vert $. The static contribution $V_{dc}^2$ has been dropped of the equation since it cannot drive resonantly the modes. Following the concept of cavity optomechanics, the mechanical resonator having the higher resonance frequency in the coupled system is chosen as the phonon-cavity. In a two-tone driving scheme, we exploit one driving tone with frequency $\Omega_{d}$  to weakly probe one of the coupled membranes around its resonance frequency ($\Omega_{1}$ or $\Omega_{2}$), and the other tone with frequency $\Omega_{p}$ to pump the phonon-cavity at its sideband $\sim\Omega_{1}\pm\Omega_{2}$. 
The $V_{ac}$ carrying the two tones therefore can be written in the form $V_{ac}\left(\Omega_p, \Omega_d\right) $ = $\frac{\mu_{p}}{2}e^{-i \Omega_pt}+\frac{\mu_{d}}{2}e^{-i \Omega_dt}+c.c$. The $\mu_{p}$ and $\mu_d$ are complex amplitudes corresponding to the $V_{ac}\left(\Omega_p\right)$ and the $V_{ac}(\Omega_d)$ components, respectively. Eq.\ref{eqn:motion} is solved in the rotating frame through looking for the mechanical displacement $X_{1(2)}(t)$=$\frac{x_{1(2)}(t)}{2}e^{-i \Omega_dt}+c.c$ mainly driven by the probe signal, and the displacement $X_{2(1)}(t)$ also depends on the interaction between the probe tone and the pump tone. The $x_{1,2}$ are the slowly varying complex amplitudes of the mechanical displacement, corresponding to each membrane's motion. In analogy with microwave optomechanics, we define the coupling strength between the phonon-cavity (SiN membrane, with index 1) and its coupled Al drum (with index 2) as $G=\frac{\partial \Omega_1}{\partial X_2}$ and the single phonon coupling strength $g_{0}$=$G\sqrt{\frac{\hbar}{2m_2\Omega_2}}$, where $\sqrt{\frac{\hbar}{2m_2\Omega_2}}$ is the zero-point fluctuations of the mechanical mode indexed 2 with resonance frequency $\Omega_2$. While the experiment is by no means anywhere close to the quantum regime, we introduce this language as a commodity for a direct comparison with the  usual formalism of optomechanics.
We first consider a two-tone driving scheme, which has been generally investigated in optomechanics. The phonon-cavity is probed with a frequency $\Omega_d$=$\Omega_1 + \delta$ and is pumped at the frequency $\Omega_p$=$\Omega_1 \pm \Omega_2 + \Delta$. The $\delta$ parameter is therefore a frequency detuning from frequency $\Omega_1$ for the probe tone. The $\Delta$ parameter is a frequency detuning for the pump, from the frequency $\Omega_1 \pm \Omega_2 $. Solving Eq.\ref{eqn:motion}, the probed mechanical displacement $x_1$ is given by
\begin{equation}
  x_1 =  \frac{f_d }{2m_1 \Omega_1} \frac{1}{\chi_1^{-1} \pm n_{p} g_0^2 \chi_2},
  \label{eqn:analog}
\end{equation}
where $n_p$=$\frac{2\vert f_p \vert ^2}{m_1\Omega_1^2}\frac{1}{\hbar \Omega_p}$ is the phonon number generated by the pumping force $f_p$=$\frac{C_{g0}V_{dc}\mu_p}{d}$ and $f_d$ = $\frac{C_{g0}V_{dc}\mu_d}{d}$ is the driving force generated by the probe signal. Both $f_d$ and $f_p$ are complex amplitudes. The $\chi_1=\frac{1}{-\delta - i \frac{\gamma_1}{2}}$ and $\chi_2=\frac{1}{\Delta-\delta-i\frac{\gamma_2}{2}}$ are susceptibilities of the phonon-cavity and the coupled electromechanical resonator. The ``-" and ``+" symbols in Eq.\ref{eqn:analog} correspond to a ``red" or ``blue" sideband pumping scheme, respectively. Similarly, the probe tone can be applied at $\Omega_d$ = $\Omega_2+\delta$ in order to measure the Al drum response, whose theoretical analyses are shown in the SI.
\begin{figure}
  \includegraphics[width=0.55\textwidth]{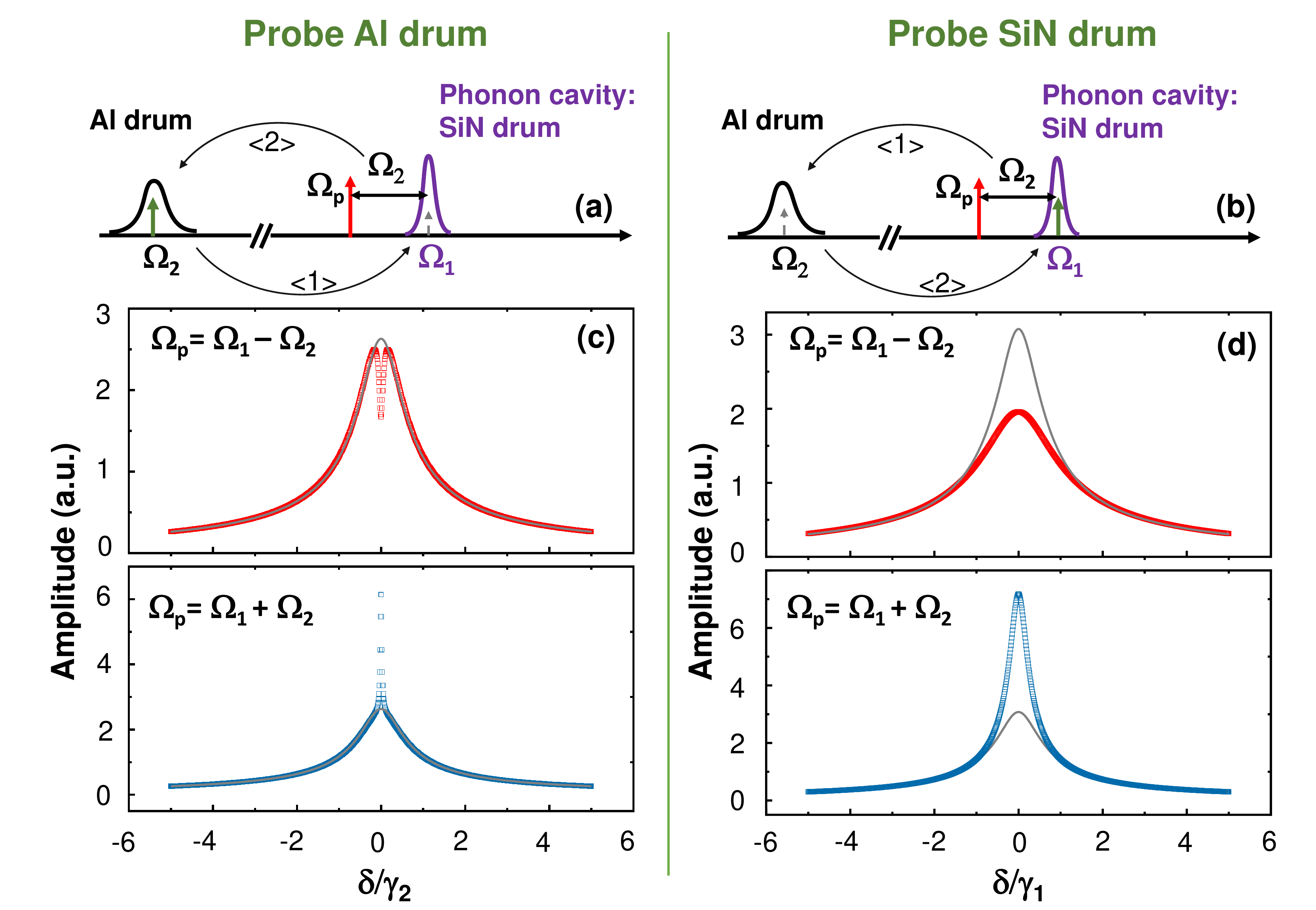}
  \caption{Diagram of the red sideband pumping scheme of the phonon-cavity SiN drum  $\Omega_p$=$\Omega_{1}$-$\Omega_{2}$+$\Delta$, while probing (a) the Al drum at frequency $\Omega_d$= $\Omega_{2}+\delta$ and (b) the SiN drum at frequency $\Omega_d$= $\Omega_{1}+\delta$. For a blue sideband pumping scheme $\Omega_p$=$\Omega_{1}$+$\Omega_{2}$+$\Delta$, the $\Omega_p$ arrow therefore is above $\Omega_{1}$, detuned by $\Omega_{2}$ (not shown). $<1>$ corresponds to the probe and the pump tone frequency up-conversion process in (a) and down-conversion process in (b). $<2>$ feed back process of the generated phonons, corresponding to the frequency down-conversion process in (a) and up-conversion process in (b). Simulated mechanical response of (c) the Al drum and (d) the SiN drum, corresponding to the red and blue sideband pumping of the phonon cavity (see $\Omega_p$ in legend). The gray curves are the mechanical responses when there is no pump tone, $n_p$=0. Both blue and red curves are computed with $n_p g_0^2$=$\gamma_{2} \gamma_{1}/7$, $f_d / (2m_{2} \Omega_{2})$= $f_d / (2m_{1} \Omega_{1})$, and $\Delta$=0.} 
  \label{sch:calcu}
\end{figure}

The probed mechanical displacement exhibits a behavior similar to optomechanically induced transparency and amplification \cite{weis2010optomechanically, hocke2012electromechanically}. In this two-tone scheme, the mechanical interaction can be decomposed into two coherent steps for phonon exchange, as indicated in the Figure \ref{sch:calcu}(a) and \ref{sch:calcu}(b), where each graph corresponds to one of our two different probing cases. In a first step, the probe and pump tone create a ``radiation pressure" force acting on the unprobed membrane that excites its mechanical vibrations, corresponding to the process $<1>$ in the Figure \ref{sch:calcu}(a) and \ref{sch:calcu}(b). Then, these generated mechanical phonons are fed back to the probed resonator, corresponding to the process $<2>$. An interference therefore is built between these phonons described by the term $n_pg_0^2\chi_2$ acting on the unprobed membrane in Eq. \ref{eqn:analog} and the initial probe signal corresponding the term $\chi_1$. This interference can be destructive or constructive, which depends on pumping the cavity at its red or blue sideband. Impacts of this interference on the resonance peaks measured with the two probe configurations are illustrated in Figure \ref{sch:calcu}(c) and \ref{sch:calcu}(d), taking taking into account the fact that the phonon cavity linewidth $\gamma_{SiN}$ is about two orders smaller than that of the coupled Al drum (our experimental conditions). As indicated in Eq.\ref{eqn:analog}, the linewidth of the unprobed mechanical resonator determines the frequency bandwidth of the transparency and amplification effects. For instance, when the phonon-cavity is red sideband pumped, the Lorentzian curve of the Al drum mechanical response exhibits a narrow dip inside its lineshape due to the fact $\gamma_{SiN} < \gamma_{Al}$, as shown in Figure \ref{sch:calcu}(c). On the contrary, the probed signal passing through the SiN drum is fully suppressed, as the $\gamma_{Al}$ determines the linewidth of the transparency window. This phenomenon is quite different from conventional optomechanical systems in which the mechanical damping rate is usually much smaller than that of the coupled cavity. These results demonstrate that both transparency and amplification windows can be controlled through engineering the mechanical damping rate in the phonon-cavity system.
\begin{figure}
  \includegraphics[width=0.55\textwidth]{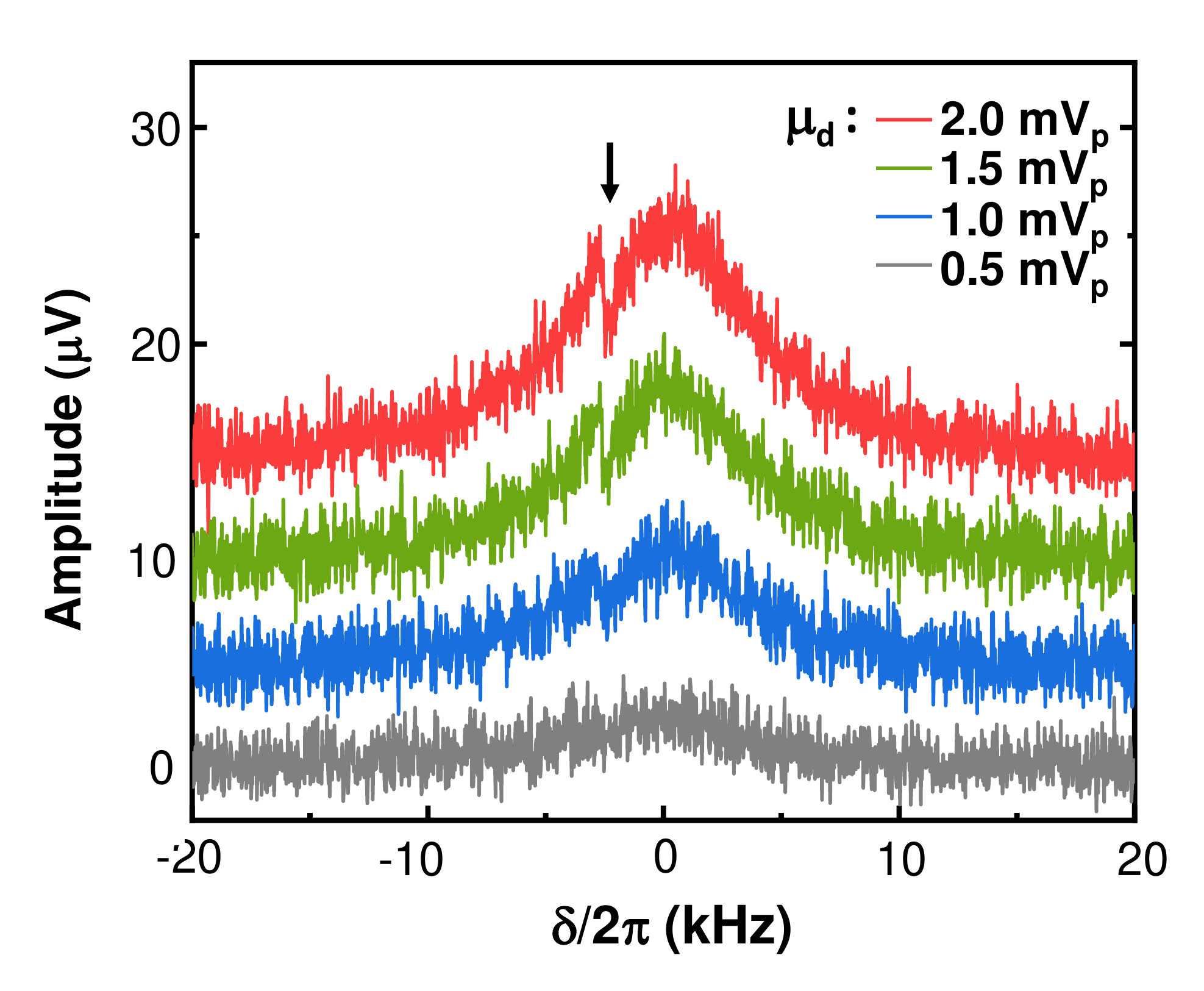}
  \caption{Mechanical response of the Al drum when the phonon-cavity is pumped at its red sideband at a frequency $\Omega_p /2 \pi = \Omega_{1} /2 \pi - \Omega_{2} /2 \pi - 2.5$ kHz with an ac pump amplitude $V_{ac}(\Omega_p)$ = 100 m$V_p$. These curves were measured with different probe ac voltages $\mu_d$ applied to the Al drum, from 0.5 mV$_p$ to 2.0 mV$_p$. The $\delta$ is the probe frequency detuning from the resonance frequency of Al drum, $\Omega_d$ = $\Omega_{2} + \delta$. When the detuning $\delta$ matches $\Delta$ = -2.5 kHz, a clear dip is visible (arrow, see text).}
  \label{sch:red_omegaX}
\end{figure}

\textbf{Electromechanically induced transparency and amplification.}  In order to build the interference process, one of the key points is that the unprobed mechanical resonator should provide enough phonons to be fed back by the pump tone, generating the interference with the initial probe tone. The motion equation Eq.\ref{eqn:motion} indicates that the energy transferred between the two movable membranes is determined by the effective pumping force $f_p \frac{X_{1(2)}(t)}{d}$, when the probe tone is driving the mechanical displacement $X_{1(2)}(t)$ around its resonance frequency $\Omega_{1(2)}$. 
Therefore, the probe tone should have a large amplitude $V_{ac}(\Omega_d)$ in order to increase the pump efficiency and provide a large number of phonons for the interference process. To demonstrate it, the phonon-cavity is red sideband pumped at a frequency $\Omega_p /2\pi$=$\Omega_{1} / 2\pi -\Omega_{2} /2\pi +\Delta$ using $\Delta$=-2.5 kHz with a fixed pump amplitude $V_{ac}(\Omega_p)$ = 100 mV$_p$ and $V_{dc}$ = 2 V. Figure \ref{sch:red_omegaX} shows the linear response of the Al drum probed by different $V_{ac}(\Omega_d)$, as a function of the frequency detuning $\delta$. At the largest drive amplitudes, a clear dip is visible when the two detunings are matched, $\delta$ = $\Delta$.
\begin{figure}
  \includegraphics[width=1.0 \textwidth]{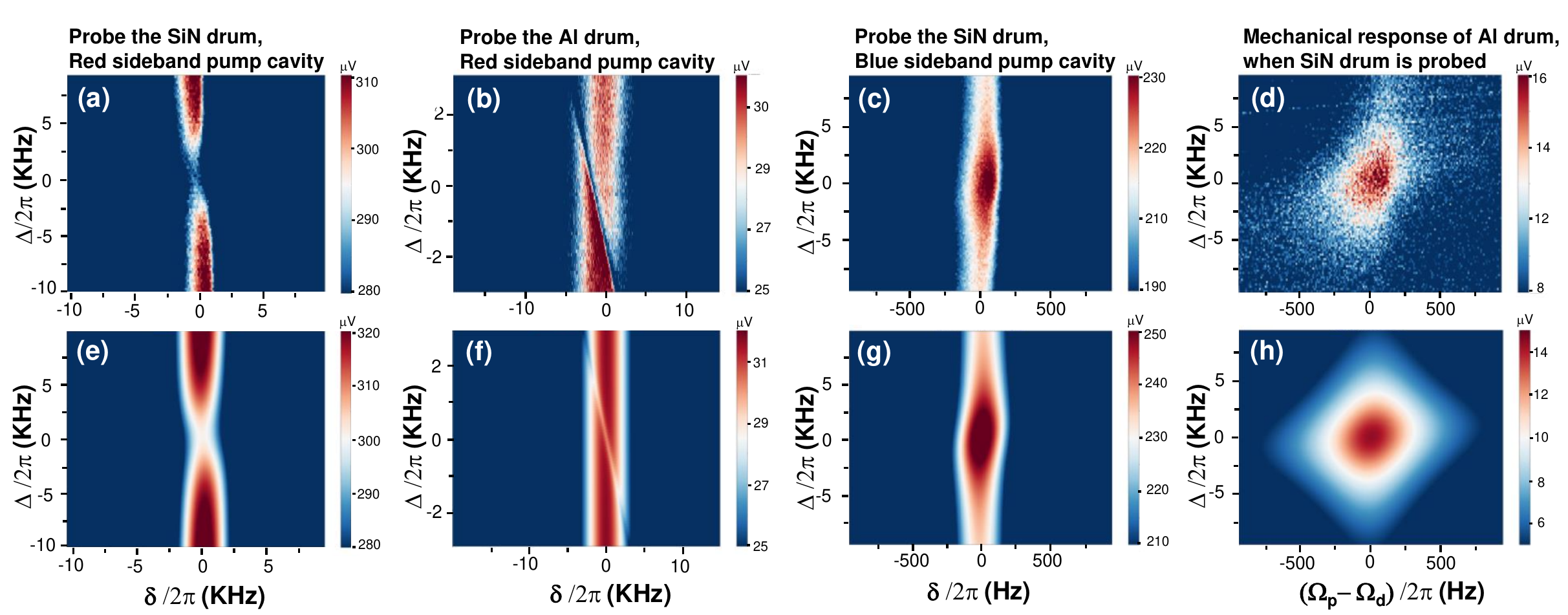}
  \caption{Electromehanically induced transparency and amplification in two capacitively coupled membranes. In a red sideband pumping scheme, (a) electromechanical response of the phonon-cavity measured with driving amplitudes $V_{dc}$ = 4 V,  $V_{ac}(\Omega_d)$ = 1 mV$_p$, $V_{ac}(\Omega_p)$ = 70 mV$_p$, and probed at $\Omega_d = \Omega_{1}+\delta$; (b) mechanical response of the Al drum measured with driving amplitudes $V_{dc}$ = 2 V,  $V_{ac}(\Omega_d)$ = 3 mV$_p$, $V_{ac}(\Omega_p)$ = 100 mV$_p$, and probed at $\Omega_d = \Omega_{2}+\delta$. In a blue sideband pumping scheme, (c) the phonon-cavity response obtained with driving amplitudes  $V_{dc}$ = 4 V, $V_{ac}(\Omega_d)$ = 0.7 mV$_p$, $V_{ac}(\Omega_p)$ = 70 mV$_p$, and probed at $\Omega_d$ = $\Omega_{1} + \delta$; (d) simultaneously measured the corresponding spectra at the frequency $\Omega_p - \Omega_d$. The spectra correspond to the process $<1>$ marked in Figure \ref{sch:calcu} (b), in a blue sideband pumping scheme. The $\Delta$ is the frequency detuning regarding the pump tone, with $\Omega_p$=$\Omega_{1} \pm \Omega_{2}$+$\Delta$ for a red or blue sideband pumping scheme. (e)-(h) Simulation results for the measurements shown in (a)-(d),  which were performed by using the theoretical model described in Eq.\ref{eqn:motion} and all experimental parameters mentioned above.} 
  \label{sch:2Dplots}
\end{figure}

Electromechanical responses of these coupled resonators have been measured as a function of the pump tone detuning $\Delta$ and the probe frequency detuning $\delta$, for both red and blue sideband pumping schemes, as shown in Figure \ref{sch:2Dplots}(a)-\ref{sch:2Dplots}(d). These measurement results clearly indicate that the linewidth of the interference window corresponding to the pump tone tuning $\Delta$ is related to the unprobed membrane. This is because the unprobed membrane in the coupled system acts as a phonon transfer station. Within the bandwidth of the unprobed resonator, phonons can be generated from the interaction between the pump and the probe tone and can coherently create constructive or destructive interferences with the probe tone. 
Figure \ref{sch:2Dplots}(c) and \ref{sch:2Dplots}(d) show the simultaneous measurement results of the electromechanical responses of the probed and the unprobed membranes, in which the phonon-cavity is driven at a frequency $\Omega_d$=$\Omega_{1}+\delta$ and is pumped at its blue sideband $\Omega_p$=$\Omega_{1}+\Omega_{2}+\Delta$. Figure \ref{sch:2Dplots}(c) clearly shows a constructive interference result as the probed signals have been amplified within the interference window and the Figure \ref{sch:2Dplots}(d) indicates phonon generation in the mechanical mode of the Al drum. The mechanical displacements of both membranes $X_{1(2)}$, corresponding to different driving and pumping schemes, have been calculated based on the capacitive coupling model described by Eq. \ref{eqn:motion}. The calculated mechanical displacements are converted into electrical signal amplitude in a microwave measurement scheme and are shown in Figure \ref{sch:2Dplots}(e)-(h), by using the relation $V_{out}$ = $\omega Z_0 C_{g0} X_{1(2)} V_{\mu w}/(2d)$ \cite{zhou2021high}. Here, $\omega$ and $V_{\mu w}$ are the frequency and the amplitude of the microwave signal for detection, respectively. The parameter $Z_0$ is the impedance of the measurement chain (here, 50 Ohm). The measurement results have been quantitatively fitted within $\sim$10$\%$ error bar by taking effective masses for the SiN membrane of $m_{1} \approx$ 4.4$\times 10^{-14}$ kg and for the Al drum of $m_{2} \approx$ 4.41$\times 10^{-13}$ kg, and by using the experimental parameters mentioned above. 

Besides, we also evaluate the maximum coupling rate $g$ = $g_0 \sqrt{n_p}$ by taking the largest pump force performed in this measurement, which has been generated by applying $V_{dc}$ = 4 V and $V_{ac}(\Omega_p)$ = 70 mV$_p$. It gives $g\approx$ 1024 rad/s , which remains smaller than
$ \gamma_{SiN}=\Omega_{SiN}/Q_{SiN}$ and $\gamma_{Al}=\Omega_{Al}/Q_{Al}$. It means that the pumping force does not provide enough phonons to drive the coupled mechanical system into the so-called strong coupling regime. In the present room temperature measurement, the pumping force is mainly limited by the large thermal background noise due to the heating effects brought in by the high pumping power. On the contrary, in the low temperature range (e.g. mK range), this limitation will be greatly suppressed. In addition, the typical damping rates of both SiN and Al mechanical resonators are $\sim$10$^2$ Hz in mK range  \cite{zhou2019chip, teufel2011sideband} with resonance frequencies in the MHz range, which could give this kind of device an access to the strong coupling regime.

\begin{figure}
  \includegraphics[width=0.55\textwidth]{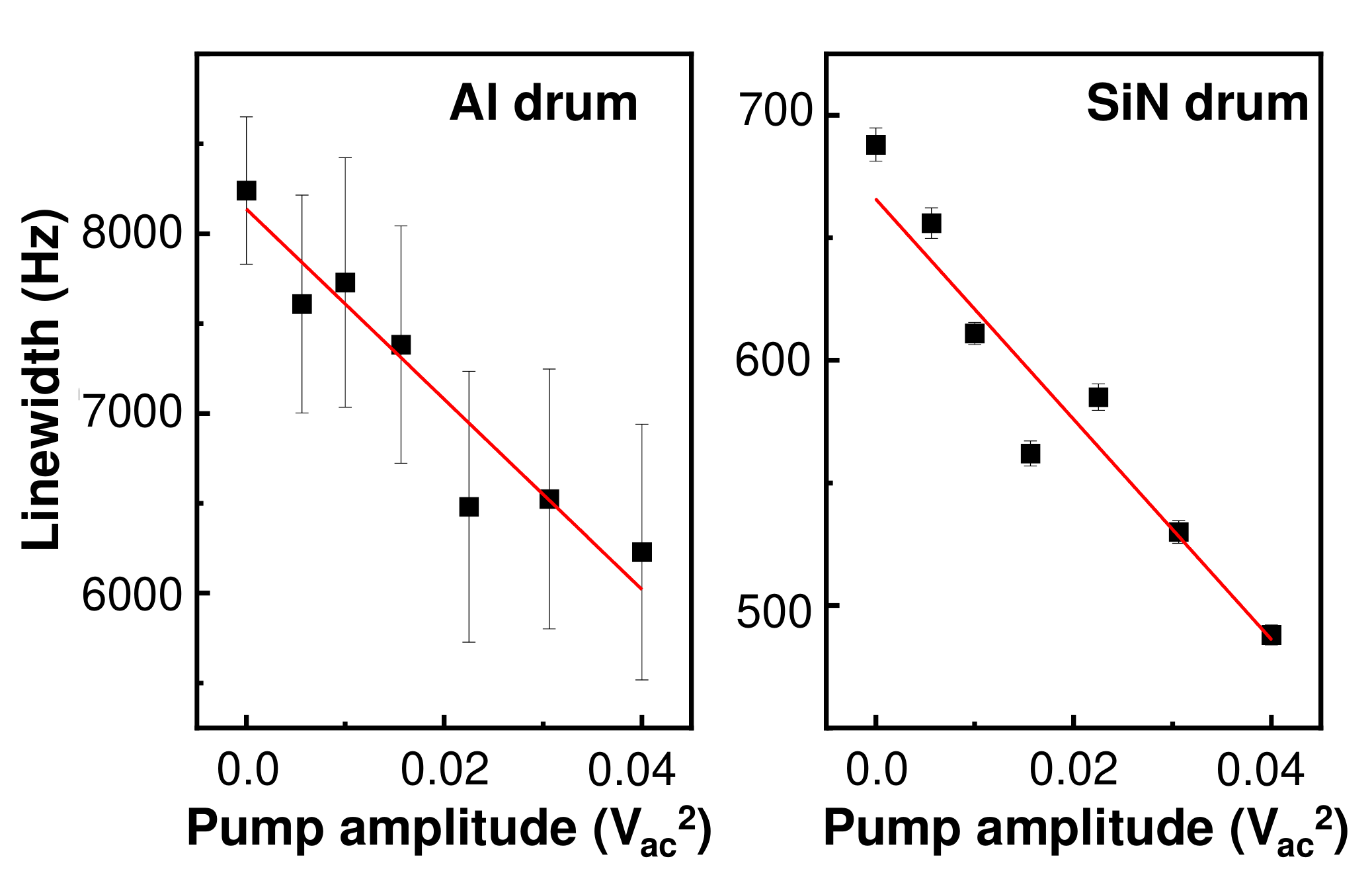}
  \caption{In a blue sideband pumping scheme $\Omega_p$ = $\Omega_{1}+\Omega_{2}$, the linewidth of the Al drum and the SiN drum are decreasing as a function of the $\textit{ac}$ pump power  [$V_{ac}^2(\Omega_p)$]. The dc bias is $V_{dc}$ = 4V, the probing voltage $V_{ac}(\Omega_d)  \approx$ 500 $\mu V$ for the Al drum and the probing voltage $V_{ac}(\Omega_d) \approx$ 200 $\mu V$ for the SiN drum. Red lines are linear fits of the data (see text).} 
  \label{sch:bluePpBW}
\end{figure}

\textbf{Backaction effects on mechanical damping rate.} In optomechanical systems, the mechanical damping rate can be modulated by the cavity backaction force, which comes from the confined electromagnetic energy generated by pumping the coupled optical or microwave cavity at its sideband. Two capacitively coupled mechanical resonators give more degrees of freedom and provide an opportunity to observe the cavity force acting on both resonators, including the one acting as a phonon-cavity. 
In order to observe this effect, we exploit a large $\textit{ac}$ signal to pump the blue sideband of the phonon-cavity, and apply a small $\textit{ac}$ one to respectively excite mechanical displacements of the SiN or Al drum to be just above the noise floor of the measurement chain. As shown in Figure \ref{sch:bluePpBW}, the linewidth of both coupled membranes decreases with increasing the pump power ($V_{ac}^2$), exhibiting the typical ``optical (anti-)damping effect"  \cite{aspelmeyer2014cavity}. We can use a single-tone driving scheme to model this effect in two capacitively coupled membranes  based on Eq. \ref{eqn:motion}, in which the phonon cavity is pumped at the frequency $\Omega_1 + \beta$. Here, the $\beta$ is a global frequency detuning, with $\beta \sim \Omega_2$ for the blue sideband and $\beta \sim -\Omega_2$ for the red sideband. The sideband scheme pumping the cavity gives rise to two satellite signals, corresponding to Stokes- and anti-Stokes scattering \cite{zhou2021electric}. The mechanical susceptibility is altered due to interactions between the pump tone and these generated Stokes and anti-Stokes processes, yielding modulations of the mechanical damping rates \cite{aspelmeyer2014cavity}. By employing a similar approach as the one applied to the electric circuit modeling of microwave optomechanics  \cite{zhou2021electric}, we therefore deduce the additional damping term $\gamma_{2opt} $ of the Al drum (with index 2) due to the back-action force of the phonon-cavity, but also the additional damping term $\gamma_{1opt}$ for the phonon-cavity (see details in SI), as shown in Eq.\ref{eqn:optdamping}. The additional damping terms make the initial mechanical damping rate $\gamma_{1(2)}$ become $\gamma_{1(2)eff}$=$\gamma_{1(2)}$+$\gamma_{1(2)opt}$,
\begin{equation}
\begin{aligned}
\gamma_{2opt} &=  \frac{ {\vert f_p \vert}^2}{4 m_1 m_2\, d^2\, \Omega_1 \Omega_2} \left[ \frac{\gamma_1}{(\Omega_2 + \beta)^2 + \frac{\gamma_1^2}{4}} - \frac{\gamma_1}{(\Omega_2 - \beta)^2 + \frac{\gamma_1^2}{4}} \right], \\
\gamma_{1opt} &=  \frac{ {\vert f_p \vert}^2}{4 m_1 m_2\, d^2\, \Omega_1 \Omega_2} \left[ \frac{\gamma_2}{(\Omega_2 - 2\Omega_1 - \beta)^2 + \frac{\gamma_2^2}{4}} - \frac{\gamma_2}{(\Omega_2 - \beta)^2 + \frac{\gamma_2^2}{4}} \right].
  \label{eqn:optdamping}
\end{aligned}
\end{equation}
In our measurement of a blue sideband pumping scheme with $\beta$=$\Omega_{2}$, Eq.\ref{eqn:optdamping} demonstrates the fact that the effective damping rate of the Al drum $\gamma_{2eff}$ is inversely proportional to the initial damping rate of the coupled phonon-cavity, $\gamma_{1}$, in accordance with the ``optical damping effect" in an optomechanical system. While conversely, for the damping rate of the phonon-cavity, $\gamma_{1eff}$ is $\propto 1/ \gamma_{2}$. If we consider the experimental condition $\Omega_{SiN}, \Omega_{Al} \gg \gamma_{SiN}$, $\gamma_{Al}$, Eq. \ref{eqn:optdamping} leads to $\gamma_{1eff}$/$\gamma_{2eff} \approx \gamma_{1} / \gamma_{2}$. In our measurement, slopes of the linewidth versus $V_{ac}^2$ can be obtained from the linear fit of the measurement results shown in Figure  \ref{sch:bluePpBW}, for both membranes. The ratio between these slopes  gives $\approx$11.8, which is in extremely good agreement with the value of $\gamma_{SiN}$/$\gamma_{Al} \approx$12.0 expected from our analytic model. 

\section{SUMMARY}
In conclusion, we have built a phonon-cavity optomechanical analogue, made of two distinct electromechanical resonators, consisting of a SiN drum and an Al drum. A simple coupling model is built to describe the mechanical motion transduction within the two capacitively coupled electromechanical resonators. In a two-tone driving scheme, electromechanically induced transparency and amplification of the injected signals have been created and manipulated at room temperature. The unique device structure allows to observe the phonon-cavity force affecting the mechanical damping rates of both coupled mechanical resonators. These optomechanical features are captured by the linear mechanical motion equations, providing theoretical analyses that quantitatively fit our measurement results. These investigations establish connections between a setup with two directly coupled movable objects and a standard optomechanical system in the classical regime. This type of capacitively coupled distinct electromechanical systems can be useful to build multimode optomechanical systems for both the classical and quantum regimes, providing a new degree of freedom in engineering photon/phonon processing, such as signal delay, storage and amplification.

\begin{acknowledgement}
X.Z. conceived the design of the experiment. A.P. collected data, H.X. participated theoretical modeling, S.V. fabricated the device. X.Z built the setup, performed the measurements, and developed the analytic model. X.Z. wrote the manuscript. All authors contributed to scientific discussions and corrections of the manuscript. Both E.C. and X.Z. supervised the project. 

We would like to acknowledge support from the STaRS-MOC project No. 181386 from Region Hauts-de-France (X.Z.), the project No. 201050 from ISITE-MOST (X.Z.), the ERC CoG grant ULT-NEMS No. 647917 (E.C.). The research leading to these results has received funding from the European Union's Horizon 2020 Research and Innovation Programme, under grant agreement No. 824109, the European Microkelvin Platform (EMP). This work was partly supported by the French Renatech network and RF-MEMS platform in IEMN. The authors acknowledge SMMiL-E, IRCL, and Contrat de Plan Etat-Région CPER Cancer 2015-2020 for technical support on wire bonding. 

\end{acknowledgement}


\begin{suppinfo}

\begin{itemize}
  \item Filename: Analytical calculation for two-tone driving scheme
  \item Filename: Analytical calculation for single-tone sideband pumping
scheme: analogy to optomechanical damping effect
\end{itemize}
%


%
\end{suppinfo}


\bibliography{achemso-demo}


\end{document}



\section{Analytical calculation for two-tone driving scheme}

This simple device structure allows to consider two parametrically coupled electromechanical resonators as a single capacitor $C_g\left(X_1, X_2\right)$ consisting of two parallel and movable membranes. The mechanical displacement of each membrane is described by $X_1(t)$ and $X_2(t)$ resonating at the frequency $\Omega_1$ and $\Omega_2$,  with $\Omega_1 > \Omega_2$. We therefore model these two coupled drums in the linear response regime, driven by an electrostatic force 
\begin{equation}
F_{1,2}(t)=\frac{(V_{dc}+V_{ac})^2}{2}\frac{\partial }{\partial X_{1,2}}C_g\left(X_1, X_2\right),
  \label{eqn:forceS}
\end{equation}
via the following equations of motion for the displacement $X_1(t)$ and $X_2(t)$,

\begin{equation}
\begin{aligned}
  \ddot{X}_1 + \gamma_1\dot{X_1} +\Omega^2_{1} X_1 &=  \frac{V_{ac}V_{dc}}{d\,m_1} C_{g0}\left[1-\frac{2(X_2-X_1)}{d}\right], \\
  \ddot{X}_2 + \gamma_2\dot{X_2} +\Omega^2_{2} X_2 &=  \frac{V_{ac}V_{dc}}{d\,m_2} C_{g0}\left[-1+\frac{2(X_2-X_1)}{d}\right].
  \label{eqn:motion}
\end{aligned}
\end{equation}
Here,  the $\gamma_{1, 2}$ is the mechanical damping rate, the $m_{1,2}$ is the effective mass, and $C_{g0}$ is the initial capacitance between two membranes separated by a distance $d$. The driving force $F_{1,2}(t)$ is modeled as a simple parallel plate capacitor, and the force is truncated at the second order Taylor expansion of the $C_g(x)$, $\approx C_{g0}(1-\frac{x}{d}+\frac{x^2}{d^2})$, with $x(t)$=$X_2(t)-X_1(t)$ In the  Eq.\ref{eqn:motion}, an approximation $2V_{dc}V_{ac}+V_{ac}^2 \approx 2V_{dc}V_{ac}$ has been made by considering a general case in measurements: $V_{dc} \gg \vert V_{ac} \vert $. The static contribution $V_{dc}^2$ has been dropped of the equation since it cannot drive resonantly the modes; note however that this term can be employed to tune the resonance frequencies \cite{zhou2021high}. We shall not refer to this possibility in the present work.

To demonstrate phonon-cavity in a two-tone driving scheme, the membrane having the higher resonance frequency is chosen as phonon-cavity (with index 1). The other coupled mechanical resonator with the lower resonance frequency is marked with index 2. We exploit one driving tone with frequency $\Omega_{d}$  to weakly probe one of the coupled membranes around $\Omega_{1}$ or $\Omega_{2}$ and the other one with frequency $\Omega_{p}$ to pump the phonon-cavity at its sideband $\sim\Omega_{1}\pm\Omega_{2}$. Therefore, we also write $V_{ac}$ in the form of $V_{ac}(\omega_p, \omega_d)$ = $\frac{\mu_{p}}{2}e^{-i \Omega_pt}+\frac{\mu_{d}}{2}e^{-i \Omega_dt}+c.c$. The Eq.\ref{eqn:motion} can be analytically solved in the rotating frame through looking for the displacement driven by the probe signal, $X_{1(2)}(t)$=$\frac{x_{1(2)}(t)}{2}e^{-i \Omega_dt}+c.c$ and the displacement of the other coupled membrane  generated by the frequency mixing between the probe and the pump signals, $X_{2(1)}(t)$=$\frac{x_{2(1)}(t)}{2}e^{-i (\Omega_p \mp \Omega_d)t}+c.c$. The $x_{1(2)}$ is the slowly varying complex amplitudes of mechanical displacements. 

First, we drive the phonon-cavity at the frequency with small amplitude around its resonance frequency $\Omega_1$ with the frequency detuning $\delta$, $\Omega_d$=$\Omega_1$+$\delta$. (a) Pump the photon-cavity at its red sideband with the frequency $\Delta$ detuned from $\Omega_{1}$-$\Omega_{2}$, $\Omega_p$=$\Omega_{1}$-$\Omega_{2}$+$\Delta$. Based on an approximation that $\Omega_1^2 - \Omega_d^2 \approx 2\Omega_1(\Omega_1-\Omega_d)$, the analytical solution of Eq.\ref{eqn:motion} gives
\begin{equation}
\begin{aligned}
  x_1 &=  \frac{f_d }{2m_1 \Omega_1} \frac{1}{\frac{1}{\chi_1}-\frac{\vert f_p \vert ^2 \chi_2}{4 m_1 m_2 d^2 \Omega_1 \Omega_2}}, \\
  x_2 &= -\frac{f_p^{*}}{2m_2 \Omega_2} \frac{x_1}{d} \chi_2 
\end{aligned}
  \label{eqn:redProb1Solu}
\end{equation}
(b) For pumping the photon-cavity at its blue sideband at the frequency $\Omega_p$=$\Omega_{1}$+$\Omega_{2}$+$\Delta$, it arrives 
\begin{equation}
\begin{aligned}
  x_1 &=  \frac{f_d }{2m_1 \Omega_1} \frac{1}{\frac{1}{\chi_1}+\frac{\vert f_p \vert ^2 \chi_2}{4 m_1 m_2 d^2 \Omega_1 \Omega_2}}, \\
  x_2^{*} &= \frac{f_p^{*}}{2m_2 \Omega_2} \frac{x_1}{d} \chi_2
\end{aligned}
  \label{eqn:blueProb1Solu}
\end{equation}
Here, we define the susceptibility of the phonon-cavity $\chi_1$ and the mechanical susceptibility $\chi_2$ corresponding to both red and blue sideband pumping the phonon-cavity. 
\begin{equation}
\begin{aligned}
  \chi_1 &= \frac{1}{-\delta-i \frac{\gamma_1}{2}}, \\
  \chi_2 &= \frac{1}{\Delta-\delta-i \frac{\gamma_2}{2}},
  \label{eqn:sus}
\end{aligned}
\end{equation}
The $f_p$ = $\frac{C_{g0}V_{dc}\mu_p}{d}$ and $f_d$ = $\frac{C_{g0}V_{dc}\mu_d}{d}$ are complex amplitudes respectively corresponding to the pumping and driving force. To have analogues of optomechanical system, we define the coupling strength as $G=\frac{\partial \Omega_1}{\partial X_2} \approx \frac{\partial \Omega_1}{\partial C_g} \frac{\partial C_g}{\partial X_2} \approx$-$\frac{\Omega_1}{2d}$. It gives single phonon coupling strength $g_{0}$=$G\sqrt{\frac{\hbar}{2m_2\Omega_2}}$, where $\sqrt{\frac{\hbar}{2m_2\Omega_2}}$ is the zero-point fluctuations of the coupled membrane with resonance frequency $\Omega_2$. Therefore, the term of $\frac{\vert f_p \vert ^2}{4 m_1 m_2 d^2 \Omega_1 \Omega_2}$ can be re-written as $n_p g_0^2$ through making a definition of the phonon number $n_p \approx \frac{2\vert f_p \vert ^2}{m_1\Omega_1^2}\frac{1}{\hbar \Omega_p}$, generated by the pump tone. Then, Eq.\ref{eqn:blueProb1Solu} becomes:
\begin{equation}
  x_1 =  \frac{f_d }{2m_1 \Omega_1} \frac{1}{\chi_1^{-1} \pm n_{p} g_0^2 \chi_2},
  \label{eqn:analog}
\end{equation}
where "-" and "+" symbols correspond to "red" and "blue" sideband pumping scheme. 

Second, we probe the coupled membrane at the frequency around its resonance frequency $\Omega_2$ with the frequency detuning $\delta$, $\Omega_d$=$\Omega_2$+$\delta$. Similarly, an approximation of  $\Omega_2^2 - \Omega_d^2 \approx 2\Omega_2(\Omega_2-\Omega_d)$ has been made. For pumping the photon-cavity at its red sideband, the analytical solution of Eq.\ref{eqn:motion} gives
\begin{equation}
\begin{aligned}
  x_2 &=  -\frac{f_d }{2m_2 \Omega_2} \frac{1}{\chi_2^{-1}-n_p g_0^2 \chi_1}, \\
  x_1 &= -\frac{f_p^{*}}{2m_1 \Omega_1} \frac{x_2}{d} \chi_1, \\
  \chi_1 &=\frac{1}{-\Delta-\delta-i \frac{\gamma_1}{2}}, \\
  \chi_2 &=\frac{1}{-\delta-i \frac{\gamma_2}{2}}.
\end{aligned}
  \label{eqn:redProb2Solu}
\end{equation}
For the blue sideband pumping, it arrives
\begin{equation}
\begin{aligned}
  x_2 &=  -\frac{f_d }{2m_2 \Omega_2} \frac{1}{\chi_2^{-1} -n_p g_0^2 \chi_1^{*}}, \\
  x_1^{*} &= -\frac{f_p^{*}}{2m_1 \Omega_1} \frac{x_2}{d} \chi_1^{*}, \\
  \chi_1^{*} &=\frac{1}{\delta-\Delta+i \frac{\gamma_1}{2}}, \\
  \chi_2 &=\frac{1}{-\delta-i \frac{\gamma_2}{2}}.
\end{aligned}
  \label{eqn:blueProb2Solu}
\end{equation}
\section{Analytical calculation for single-tone sideband pumping scheme:  analogy to optomechanical damping effect}
Here, we define the phonon-cavity is sideband pumped at $\Omega_p = \Omega_1 + \beta$, where $\beta$ is the frequency detuning from the resonance frequency of the phonon-cavity $\Omega_1$. The mechanical displacement of the coupled membrane (with the index 2) is written as $x_2(t) = \frac{1}{2} \delta x_2(t)e^{-i\Omega_2 t} + c.c$, where the complex amplitude of $\delta x_2(t)$ is the Brownian motion of the the membrane(2) \cite{zhou2021electric}. The terms where motion $x_2(t)$ multiplies pump amplitude in Eq.\ref{eqn:motion} generate harmonics at $\Omega_n = \Omega_p + n \Omega_2$, with n $\in \mathbb{Z}$. The solution can be found in the form of the  \textit{ansatz}, 
\begin{equation}
x(t) = \sum_{n=-\infty}^{+\infty} \frac{\delta x(t)}{2}e^{-i(\Omega_p + n \Omega_2)t} + c.c.
\end{equation}
In this work, we are interested only  in schemes of $n = \pm 1$, corresponding to the "down" and "up" sideband of the pump signals at the frequency $\Omega_{-} = \Omega_p  - \Omega_2 $ and $\Omega_{+} = \Omega_p + \Omega_2 $. The solution of the phonon-cavity motion equation in Eq.\ref{eqn:motion} is given by
\begin{equation}
\begin{aligned}
  x_{-} &=  \frac{e^{-i\Omega_{-}t}}{2} \frac{f_p \delta x_2^{*}}{m_1d} \frac{1}{\Omega_1^2-\Omega_{-}^2 - i \Omega_{-}\gamma_1} + c.c, \\
  x_{+} &= \frac{e^{-i\Omega_{+}t}}{2} \frac{f_p \delta x_2}{m_1d} \frac{1}{\Omega_1^2-\Omega_{+}^2 - i \Omega_{+}\gamma_1} + c.c.
\end{aligned}
  \label{eqn:plusm}
\end{equation}
It yields an extra force,  $f_{cav}= \frac{f_p}{d} x_{-}^{*} + \frac{f_p^*}{d} x_{+} + c.c$, biasing on the membrane(2), which modifies the initial mechanical susceptibility to become
\begin{equation}
\begin{aligned}
\chi_2(\Omega) &= \frac{1}{2m_2 \Omega_2} \frac{1}{ (-\Omega - i \frac{\gamma_2}{2}) + \Sigma}, \\
\Sigma &= \frac{ {\vert f_p \vert}^2}{4 m_1 m_2 d^2 \Omega_1 \Omega_2} ( \frac{\Omega_2 + \beta - i \frac{\gamma_1}{2}}{(\Omega_2 + \beta)^2 + \frac{\gamma_1^2}{4}}-\frac{\Omega_2 - \beta -i\frac{\gamma_1}{2}}{(\Omega_2-\beta)^2 +\frac{\gamma_1^2}{4}} )
\end{aligned}
  \label{eqn:plusm2}
\end{equation}
The imaginary part of $\Sigma$ modifies the mechanical damping rate $\gamma_2$, yielding additional damping $\gamma_{opt}$
\begin{equation}
\gamma_{opt} = n_{p} g_0^2 \left[ \frac{\gamma_1}{(\Omega_2 + \beta)^2 + \frac{\gamma_1^2}{4}} - \frac{\gamma_1}{(\Omega_2 - \beta)^2 + \frac{\gamma_1^2}{4}} \right].
\label{eqn:effdamping}
\end{equation}
The real part of $\Sigma$ contributes to a frequency shift of $\Omega_2$, 
\begin{equation}
\delta \Omega_2 = -n_{p} g_0^2 \left[ \frac{\beta +\Omega_2}{(\Omega_2 + \beta)^2 + \frac{\gamma_1^2}{4}} - \frac{\Omega_2 - \beta}{(\Omega_2 - \beta)^2 + \frac{\gamma_1^2}{4}} \right].
\label{eqn:effres}
\end{equation}
Both expressions of Eq.\ref{eqn:effdamping} and Eq.\ref{eqn:effres} refer to  "optical damping effect" and "optical spring effect" in optomechanics. The $f_{cav}$ corresponds to the cavity force, originating from electromagnetic field confined in this capacitor, consisting of two capacitively coupled membranes. 

Using same method, we could get the dynamical backaction effects on the phonon-cavity, 
\begin{equation}
\begin{aligned}
\gamma_{opt (cavity)} &= n_{p} g_0^2 \left[ \frac{\gamma_2}{(\Omega_2 - 2\Omega_1 - \beta)^2 + \frac{\gamma_2^2}{4}} - \frac{\gamma_2}{(\Omega_2 - \beta)^2 + \frac{\gamma_2^2}{4}} \right] , \\
\delta \Omega_1 &= n_{p} g_0^2 \left[ \frac{\Omega_2 -2\Omega_1 - \beta}{(\Omega_2 -2\Omega_1 - \beta)^2 + \frac{\gamma_2^2}{4}} + \frac{\Omega_2 - \beta}{(\Omega_2 - \beta)^2 + \frac{\gamma_2^2}{4}} \right].
\end{aligned}
  \label{eqn:backaction_cav}
\end{equation}
%
%
\begin{figure}
  \includegraphics[width=0.0\textwidth]{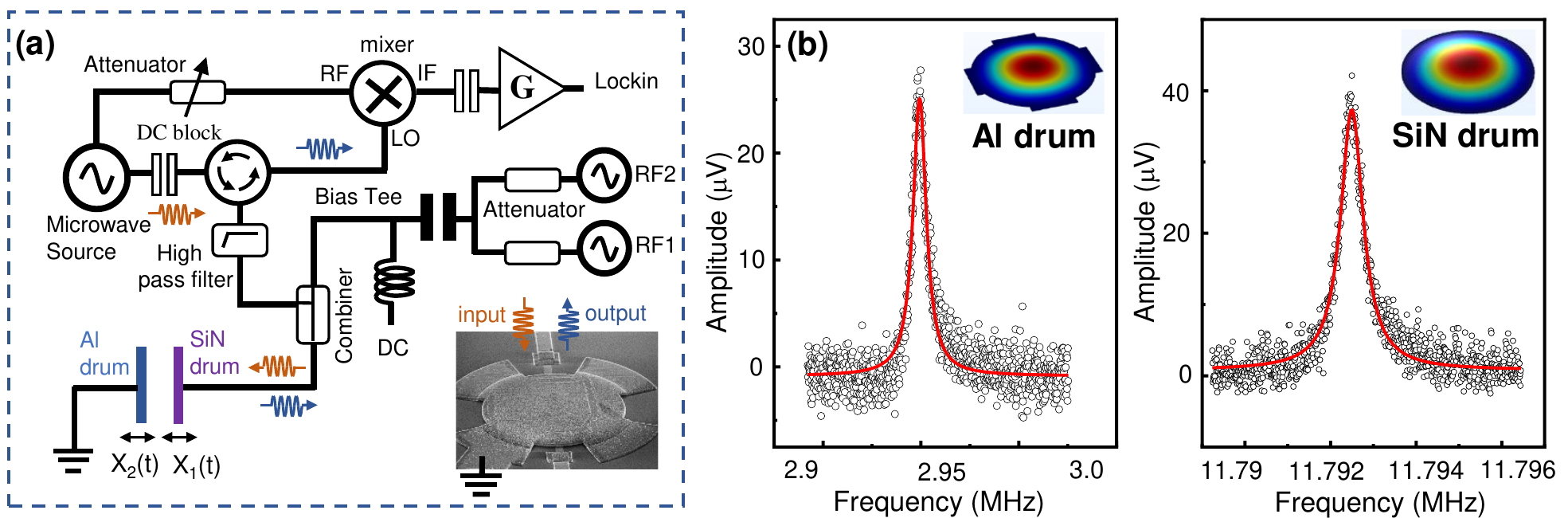}

  \label{sch:nonliear}
\end{figure}
\bibliography{supplement}